\documentclass[twocolumn,prl,aps,showpacs,superscriptaddress]{revtex4}

\begin{document}
\title{Path Distinguishability in Double Scattering of Light by Atoms}

\author{Christian Miniatura}
\affiliation{Department of Physics, Faculty of Science, %
National University of Singapore, Singapore 117542, Singapore}
\affiliation{\mbox{Institut Non Lin{\'e}aire de Nice, UMR 6618 du CNRS, %
UNSA, 1361 route des Lucioles, 06560 Valbonne, France}}
\author{Cord A. M\"uller}
\affiliation{Physikalisches Institut, Universit\"at Bayreuth, %
95440 Bayreuth, Germany}
\author{Yin Lu}
\author{Guangquan Wang}
\author{Berthold-Georg Englert}
\affiliation{Department of Physics, Faculty of Science, %
National University of Singapore, Singapore 117542, Singapore}

\date{15 April 2007}

\begin{abstract}
Wave-particle duality finds a natural application for electrons or light
propagating in disordered media where coherent corrections to transport are
given by two-wave interference. 
For scatterers with internal degrees of freedom, these corrections are
observed to be much smaller than would be expected for structureless
scatterers.
By examining the basic example of the
scattering of one photon by two spin-$\frac{1}{2}$ atoms---a case-study for
coherent backscattering---we demonstrate that
the loss of interference strength is associated with which-path information
stored by the scattering atoms.
\end{abstract}


\maketitle

Einstein's and de~Broglie's wave-particle duality (WPD)---the ability of a
quantum system to display the seemingly contradictory attributes that one
would have regarded as wave-like and particle-like and, therefore, as mutually
exclusive in pre-quantum physics---is arguably the most important
phenomenological consequence of Bohr's principle of complementarity. 
The quantitative aspects of WPD are particularly well understood 
\cite{EnglertBergou2000} in the context of two-paths interferometers where a
definite path is particle-like and the interference between the amplitudes of
the two paths is wave-like.

The wave-like interference strength is quantified by the familiar Michelson's
fringe visibility $\mathcal{V}$, and the particle-like path
knowledge is measured by the path distinguishability $\mathcal{D}$, which is
perhaps less familiar and has this operational meaning: The odds for guessing
the path right are $(1+\mathcal{D})/2$.   
The extreme situations of no path knowledge and high fringe visibility
(ideally $\mathcal{D}=0$, $\mathcal{V}=1$) or full path knowledge and no
fringes ($\mathcal{D}=1$, $\mathcal{V}=0$) are standard textbook fare. 
A rather recent experiment with a bearing on the matter discussed below is the
one carried out by Eichmann \textit{et al.} in 1993 \cite{EichmannItano}. 

The compromises allowed by the laws of physics in intermediary situations are
restricted  by the duality relation \cite{JSV,Englert96}
\begin{equation}
\label{duality}
\mathcal{D}^2 + \mathcal{V}^2 \leq 1\,.
\end{equation}
It is worth noting that well pronounced wave-like and particle-like aspects
can coexist: 
With odds for guessing the path right of 99\% ($\mathcal{D}=0.98$), we can
have well visible fringes of 20\% visibility ($\mathcal{V}=0.2$).

The two pioneering experiments that tested the duality relation employed 
two-path interferometers for atoms \cite{Rempe98} and photons \cite{Kwiat99}, 
whereby internal degrees of freedom of the interfering objects themselves were
used to provide the path information.
By contrast, in the situation that we examine here---coherent double
scattering---the which-path information is stored in the deflecting elements
of the interferometer (atoms) and not carried by the interfering objects
(photons).  

We wish to show how these notions of path knowledge and interference strength
are naturally applied to coherent wave transport. 
In the semi-classical regime of weak localization, most of the coherent
effects in wave transport can be explained by \emph{two-wave}
interference between amplitudes propagating in opposite direction
along loop-like scattering paths \cite{Bergmann84, Houches94}.
These interference corrections to transport can dramatically alter
the diffusion process and even suppress it \cite{Lagendijk,Maret}. 
They are sensitive to several ``dephasing'' processes but
most of them are circumvented at sufficiently low temperatures
\cite{Imry, Saclay, Dynamical}. 

Here we address an intrinsic dephasing mechanism which survives at zero
temperature: the path knowledge stored in the internal degrees of freedom of
the scatterers. 
Indeed, when the scatterers have an internal structure, the interference
corrections to transport are observed to be rather small, for example in the
scattering of electrons by magnetic impurities at very low temperatures 
\cite{Saclay}.

The same effect has been observed in the coherent backscattering
(CBS) of light by cold rubidium atoms \cite{Labeyrie99,Jonckheere, Mueller}. 
This coherent multiple scattering effect arises when an optically thick sample
of scatterers is illuminated by coherent light. 
It, too, results from interference of light amplitudes, here of the two
amplitudes associated with traversing the same path in opposite direction.
The endpoints of each scattering path thus play the role of Young
slits and give rise to an angular fringe pattern in the far-field.
Owing to the varying separation between the endpoints, 
these patterns have different fringe spacings but 
they \emph{all} display a bright fringe at backscattering.
Thus, the sum of all fringe patterns displays an angular peak
around the backscattering direction \cite{Cbs}.

We quantify the strength of this interference in a natural manner by 
the relative excess of the peak intensity over the background, 
the analog of the fringe visibility $\mathcal{V}$ in this context. 
For atoms with a spin-0 ground state, this CBS peak-to-background ratio 
reaches its maximal possible value of $2$ in the helicity-preserving
polarization channel \cite{Sr}, corresponding to ${\mathcal{V}=1}$, 
whereas it is very small for atoms with a degenerate
ground state \cite{Labeyrie99}, as is the situation with a Zeeman fine
structure or a hyperfine structure. 

Our point is that these results can be recast and understood,
both qualitatively and quantitatively, in terms of wave-particle duality. 
Indeed, when scattering the photon the atoms may undergo a change in their
ground state---a circumstance equally crucial in the single-scattering
situation of the Eichmann \textit{et al.\/} experiment \cite{EichmannItano}.
This is to say that the atoms can store which-path information so that the
experimenter can find out, in principle if not in practice, which of the
two atoms scattered first and which second.
As a consequence, the strength of the coherent corrections to
transport is bounded by the distinguishability of the paths inside
the sample, and the height of the CBS peak is limited by the amount of
path knowledge available.

We consider the simplest possible scenario that exhibits the effect:
double scattering off two identical spin-$\half$ atoms (atom 1 and atom 2), 
with the photon resonant with a $\half\leftrightarrow\half$ dipole
transition. In fact, this is the situation of the Eichmann \textit{et al.\/}
experiment where the scatterers are $\mathrm{Hg}^+$ ions.
This geometry is simplest for multiple scattering to set in and 
is at the heart of the CBS phenomenon.
Since our focus is on the influence of the internal atomic structure, we assume
that the atoms are so stiffly trapped that there is no relevant contribution
from the atomic recoil (the storage of CBS which-path information in the
center-of-mass degrees of freedom of mobile atoms is studied in 
Ref.~\cite{Wickles}).
Put differently, we take for granted that the atomic center-of-mass degrees of
freedom do not store which-path information. 

\begin{figure}[t]
\centerline{\includegraphics{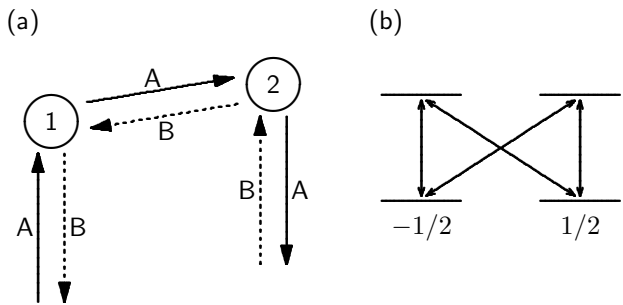}}
\caption{\label{fig:1}%
(a) The two paths in coherent backscattering. Along path A the photon
is first scattered by atom~1, then by atom~2; along path B the order is
reversed. (b) Level scheme of the ${\half\leftrightarrow\half}$
transition. Both the ground state and the excited state are doublets with
total angular momentum $j=\half$, and the magnetic quantum numbers
$m=\pm\half$ label the sublevels.}
\end{figure}

To simplify the problem further, we assume that the distance 
between the atoms is sufficiently large for the double scattering
contribution to dominate over all other multiple scattering
processes (triple, quadruple, \dots). 
As illustrated in Fig.~\ref{fig:1}, 
path A is the case when atom~1 scatters first and atom~2 second (sequence
${1\to2}$); path B is the sequence ${2 \to 1}$. 
The paths are geometrically identical but traversed in opposite directions. 
The two atoms together compose the path detector: 
the change of their internal states bears witness of the actual path. 

Since the ground states are spin-$\half$ states, the path detector is a qubit
pair, which is a 4-state system. 
During the scattering process, however, the excited states of the atoms are
involved as well, and the details of the scattering interaction determine the
over-all effect on the atoms.
This net before-to-after change in the combined ground states of both atoms is
given by a completely positive two-qubit map.

We establish this map by first recalling that the atom-photon interaction is
described by quasi-resonant point-dipole elastic scattering. 
The corresponding transition operator is proportional to
$\mathcal{T} = (\boldsymbol{d} \, \boldsymbol{d}) \otimes
|\boldsymbol{r}\rangle\langle \boldsymbol{r}|$ where $\boldsymbol{r}$ is the
atom's position vector and $\boldsymbol{d}$ is the dipole vector operator 
of the atomic transition.
The omitted proportionality factor depends on the oscillator strength of the
transition.
It determines the probability of the double scattering event and is,
therefore, crucial for an actual experiment.
But in the present context this probability is irrelevant because \emph{the
  final two-atom state is conditioned on successful double scattering}.
Bearing this conditioning in mind, we consistently leave all further
proportionality factors implicit. 

For an incoming photon with wave
vector $\boldsymbol{k}$ and transverse polarization
$\boldsymbol{\epsilon}$, the matrix elements of $\mathcal{T}$ are
\begin{equation}\label{scatensor} 
\langle m', \boldsymbol{k}' \boldsymbol{\epsilon}'
|\mathcal{T}| m, \boldsymbol{k} \boldsymbol{\epsilon} \rangle 
= \langle m', \boldsymbol{\epsilon}' | (\boldsymbol{d} \, \boldsymbol{d}) | m,
\boldsymbol{\epsilon} \rangle \ 
e^{i(\boldsymbol{k}-\boldsymbol{k}')\cdot\boldsymbol{r}}\,,
\end{equation}
where $\boldsymbol{k}'$ and $\boldsymbol{\epsilon}'$ are the wave vector and
polarization of the outgoing photon, and $m$ and $m'$ are the magnetic quantum
numbers of the initial and final ground state, respectively \cite{Mueller}.
Since the scattering is elastic, we have $|\boldsymbol{k}|=|\boldsymbol{k}'|$.

The dyadic operator
$T=(\boldsymbol{d} \, \boldsymbol{d})$ acts on the internal degrees of
freedom of the photon (the polarization states) and of the atom
(the magnetic sublevels of the angular momentum multiplets). 
Its matrix elements read
\begin{equation}
\label{intscatensor} \langle m', \boldsymbol{\epsilon}' |(\boldsymbol{d}
\, \boldsymbol{d})| m, \boldsymbol{\epsilon} \rangle 
= \langle m'|(\boldsymbol{\epsilon}'^*\cdot\boldsymbol{d})
(\boldsymbol{d}\cdot\boldsymbol{\epsilon}) | m \rangle\,.
\end{equation}
The matrix elements of the vector operator $\boldsymbol{d}$ are the
Clebsch--Gordan coefficients that characterize the coupling 
of spin-1 (photon) with spin-$\half$ (ground state) to give spin-$\half$
(excited state); all the coefficients have equal magnitude for such a
$\half\leftrightarrow\half$ transition.
As a consequence, we have effectively 
$T = (\boldsymbol{\sigma} \,\boldsymbol{\sigma})$ for initial and final ground
states, where $\boldsymbol{\sigma}$ is the Pauli
vector operator for the spin-$\half$ ground state.~\cite{threehalf}

We consider the exact backscattering geometry where
$\boldsymbol{k}=-\boldsymbol{k}'$, which we choose parallel 
to the $z$ axis of the
coordinate system; the magnetic quantum numbers $\pm1/2$ in Fig.~\ref{fig:1}(b)
also refer to the $z$ direction.  
For path A, the double scattering operator that acts on the two atomic
ground-state qubits is
\begin{eqnarray}\label{eq:TA}
T_\mathrm{A} &=& \boldsymbol{\epsilon}'^*\cdot(\boldsymbol{\sigma}_2
\boldsymbol{\sigma}_2)\cdot (\mathbf{1}-\boldsymbol{n}\boldsymbol{n})
\cdot(\boldsymbol{\sigma}_1 
\boldsymbol{\sigma}_1)\cdot\boldsymbol{\epsilon} \nonumber\\
&=&-\boldsymbol{\epsilon}'^*\cdot\boldsymbol{\sigma}_2
\,(\boldsymbol{\sigma}_2\times\boldsymbol{n})\cdot
(\boldsymbol{n}\times\boldsymbol{\sigma}_1)\,
\boldsymbol{\sigma}_1\cdot\boldsymbol{\epsilon}\,,
\end{eqnarray}
where $\mathbf{1}-\boldsymbol{n}\boldsymbol{n}$ 
is the dyadic projector onto the plane
orthogonal to the unit vector $\boldsymbol{n}$ that points from one
scatterer to the other.
The double-scattering operator $T_\mathrm{B}$ for path B is obtained by
interchanging $1\leftrightarrow 2$ in \eqref{eq:TA}. 

With $\rho_\mathrm{in}$ denoting the initial two-qubit state of the two
scattering atoms, the final states are then given by 
\begin{eqnarray}\label{eq:finrho}
\rho_\mathrm{A,B}& =& \frac{T_\mathrm{A,B} \,
\rho_\mathrm{in}\, {T_\mathrm{A,B}}^\dagger}{w_\mathrm{A,B}}\nonumber\\
\textrm{with}\quad w_\mathrm{A,B}&=&\tr{T_\mathrm{A,B} \,
\rho_\mathrm{in}\, {T_\mathrm{A,B}}^\dagger}\,,
\end{eqnarray}
where the normalizing denominators take care of all the proportionality
factors that we left implicit.
The weights of the two paths are 
$w_\mathrm{A}/(w_\mathrm{A}+w_\mathrm{B})$ and
$w_\mathrm{B}/(w_\mathrm{A}+w_\mathrm{B})$, respectively.
In addition to the initial two-atom state $\rho_\mathrm{in}$, these weights
and the final states depend on the pre-selected polarization 
$\boldsymbol{\epsilon}$ of the
incoming photon and the post-selected polarization $\boldsymbol{\epsilon}'$ of
the outgoing photon, on which the ensemble of events is conditioned. 

Since the final states of the atoms are different for the two paths, there is
which-path information stored in the atoms, which---in principle---can be
extracted by a suitable measurement, although in practice it could be very
difficult to implement such a measurement. 
The optimal measurement would provide as much path knowledge as is available,
quantified  by the distinguishability of the paths, which is given by 
\cite{JSV,Englert96,EnglertBergou2000}
\begin{eqnarray}
  \label{eq:D}
  \mathcal{D}&=&
  \frac{\tr{\bigl|w_\mathrm{A}\rho_\mathrm{A}-w_\mathrm{B}\rho_\mathrm{B}\bigr|}}
                   {w_\mathrm{A}+w_\mathrm{B}}\nonumber\\
     &=&\frac{\tr{\bigl|{T_\mathrm{A}}\,\rho_\mathrm{in}\,{T_\mathrm{A}}^\dagger
              -T_\mathrm{B} \, \rho_\mathrm{in}\,{T_\mathrm{B}}^\dagger\bigr|}}
                   {w_\mathrm{A}+w_\mathrm{B}}
\,.
\end{eqnarray}
This is supplemented by the visibility 
\begin{equation}
  \label{eq:V}
  \mathcal{V}=\frac{2\Bigl|\tr{T_\mathrm{A} \,
                     \rho_\mathrm{in}\, {T_\mathrm{B}}^\dagger}\Bigr|}
               {w_\mathrm{A}+w_\mathrm{B}}\,,
\end{equation}
the quantitative measure for the interference strength of the two paths.
Irrespective of the detailed form of $\rho_\mathrm{in}$ and the operators
$T_\mathrm{A},T_\mathrm{B}$, the duality relation \eqref{duality} is obeyed by
this distinguishability and visibility \cite{EnglertBergou2000}. 

We now restrict the discussion to symmetric initial two-qubit states
of the form $\rho_\mathrm{in}=\frac{1}{4}(\openone-%
p\boldsymbol{\sigma}_1\cdot\boldsymbol{\sigma}_2)$
with $-\frac{1}{3}\leq p\leq1$ as required by the positivity of 
$\rho_\mathrm{in}$.
This one-parameter family of initial states encompasses some cases of
particular physical interest: the completely mixed state ($p=0$);
the projector on the singlet state of vanishing total angular momentum 
($p=1$); the projector on the triplet sector of unit
total angular momentum  ($p=-\frac{1}{3}$). 
It is worth recalling that two-qubit states of this form are separable for
$p\leq\frac{1}{3}$ and entangled for $p>\frac{1}{3}$ but, as illustrated by
Eqs.~\eqref{eq:perpgeo} below, 
nothing remarkable happens to $\mathcal{D}$ and $\mathcal{V}$ at the
transition from $p<\frac{1}{3}$ to $p>\frac{1}{3}$. 

For all values of $p$, the initial state $\rho_\mathrm{in}$ 
is invariant under the interchange $1\leftrightarrow 2$ and, therefore, 
the interferometer is symmetric in the sense that both paths occur with equal
\textit{a priori} probability ($w_\mathrm{A}=w_\mathrm{B}$). 
As a consequence, the difference of operators in \eqref{eq:D} is
\emph{antisymmetric} under $1\leftrightarrow2$ and thus of the form
\begin{equation}
  \label{eq:delta}
     \frac{T_\mathrm{A} \, \rho_\mathrm{in}\, {T_\mathrm{A}}^\dagger
           -T_\mathrm{B} \, \rho_\mathrm{in}\, {T_\mathrm{B}}^\dagger}
          {w_\mathrm{A}+w_\mathrm{B}}
  =\boldsymbol{a}\cdot(\boldsymbol{\sigma}_1-\boldsymbol{\sigma}_2) 
   + \boldsymbol{b}\cdot(\boldsymbol{\sigma}_1 \times\boldsymbol{\sigma}_2)
\end{equation}
with two numerical vectors $\boldsymbol{a}$ and $\boldsymbol{b}$ that depend
on the photon polarizations $\boldsymbol{\epsilon},\boldsymbol{\epsilon}'$,
the unit vector $\boldsymbol{n}$, and the initial-state parameter $p$.
The right-hand side in \eqref{eq:delta} is a rank-2 operator with its
nonzero eigenvalues given by $\pm2\sqrt{\boldsymbol{a}^2+\boldsymbol{b}^2}$,
and so we get $\mathcal{D} = 4\sqrt{\boldsymbol{a}^2+\boldsymbol{b}^2}$
for the distinguishability of the paths.

In this manner we arrive at explicit 
expressions for $\mathcal{D}$ and $\mathcal{V}$ \cite{ISM,HYP}.
We will report the full technical details elsewhere and focus here on the
particular situation in which the line connecting the two atoms in
Fig.~\ref{fig:1}(a) is perpendicular to the incoming and outgoing propagation
directions, that is: we choose the unit vector $\boldsymbol{n}$ along the
$x$ axis.
For this perpendicular geometry, one has
\begin{eqnarray}
  \label{eq:perpgeo}
  \mathcal{D}&=&\frac{1+p+2pu}{2(1+pu)}\sqrt{1-{u'}^2}\,,
\nonumber\\
  \mathcal{V}&=&\frac{\bigl|(1+p)(1+uu')-2p(1-u')\bigr|}{2(1+pu)}\,,
\end{eqnarray}
where $u$ and $u'$ are the $x$ components of the Stokes vectors associated
with the incoming and outgoing photon polarizations \cite{BornWolf}.
These are such that
\begin{equation}
  \label{eq:border}
  \mathcal{D}\leq\left\{
    \begin{array}{l@{\ \textrm{for}\ } l}
      \sqrt{(1-\mathcal{V})\mathcal{V}} & p\leq0\,,\\[1ex]
      \displaystyle 
      \sqrt{(1-\mathcal{V})\bigl(2p/(1+p)+\mathcal{V}\bigr)} & p\geq0\,.
    \end{array}\right.
\end{equation}
Clearly, the duality relation \eqref{duality} is obeyed for all $p$ values.
The relation is only saturated for $p=1$, in which case the initial atomic
state is pure and the equal sign is expected to hold in \eqref{duality} on
general grounds \cite{Englert96}. 

For the completely-mixed initial state ($p=0$) we have
\begin{equation}
  \label{eq:chaos}
  \mathcal{D}=\half\sqrt{1-{u'}^2}\,,\qquad
  \mathcal{V}=\half(1+uu')\,.
\end{equation}
Here, the distinguishability does not depend at all on the initial
polarization---a surprising feature that is particular 
to the $\half\leftrightarrow\half$
transition in the perpendicular geometry and is not generic. 
This observation about the perpendicular geometry can be understood as follows.

Since all Clebsch--Gordan coefficients are of equal size, the first scatterer
has uniform \emph{a priori} probability of reaching either one of its ground
states, irrespective of the polarization of the incoming photon.
Yet, when conditioned on the direction into which the photon is re-emitted,
the probability is not uniform as a rule, 
but it \emph{is} for the perpendicular geometry. 
Therefore, there is no which-path information stored in the final
completely-mixed state of the atom that scatters first.

When the observed polarization of the outgoing photon is an equal-weight
superposition of the in-plane and out-of-plane linear polarizations
($u'=0$), the final state of the second scatterer is a corresponding pure
state. 
So, when finding only one atom in this pure state, we can infer the path with
certainty, but if both atoms are found in this state, we know nothing about
the path and will guess wrong half of the time. 
Both cases are equally probable, so that our betting odds are 75\%, which is
consistent with $\mathcal{D}=\half$ for $u'=0$ in \eqref{eq:chaos}, as it
should be.
In this case the visibility is $\mathcal{V}=\half$ irrespective of the
incoming polarization. 

The distinguishability is zero for an outgoing photon linearly polarized 
in the plane of the drawing in
Fig.~\ref{fig:1}(a), when $u'=1$, or perpendicular to this plane ($u'=-1$). 
The corresponding visibility takes on any value between $0$ and $1$. 
The case $\mathcal{V}=0$ happens for photons with perpendicular polarizations, 
one with in-plane polarization, the other perpendicular ($uu'=-1$). 
The case  $\mathcal{V}=1$ occurs for photons with
parallel linear polarizations, both in the plane or both perpendicular to it
($uu'=1$). 

Let us now turn to the situation of an initial singlet state ($p=1$). 
As noted above, the duality relation \eqref{duality} is then 
saturated and we have
\begin{equation}
  \label{eq:singlet}
  \mathcal{D}=\sqrt{1-{u'}^2}\,,\qquad
  \mathcal{V}=\bigl|u'\bigr|\,.
\end{equation}
The fact that the distinguishability does not depend on the initial
polarization can be understood by an argument similar to the one given above 
for the $p=0$ case. 
Irrespective of the incoming photon polarization and
for, say, a left-circular outgoing photon, the final states of the atoms
have $(m_1,m_2)=(\half,-\half)$ for path A and $(-\half,\half)$
for path B whereas the reversed situation occurs for a right-circular 
outgoing photon. 
This means that if the outgoing light is analyzed in the circular
channels ($u'=0$), perfect path knowledge is available
($\mathcal{D}=1$) and no interference will be observed ($\mathcal{V}=0$).
Conversely, if the outgoing light is analyzed in the 
channels of linear in-plane and out-of-plane polarization 
($\bigl|u'\bigr|=1$), no path knowledge is available
($\mathcal{D}=0$) and one recovers full interference strength
($\mathcal{V}=1$). 

Finally, one can think of mimicking the physics of the CBS
phenomenon by an angular average over the direction $\boldsymbol{n}$. 
We first calculate the average of the difference operator in \eqref{eq:delta}
and then compute the resulting distinguishability as the trace of its modulus.
The corresponding visibility is obtained as the angular average of the
$\boldsymbol{n}$ dependent visibility \eqref{eq:V}. 
For $p=0$, which applies to most of the available experimental
data, the largest averaged distinguishability is $\mathcal{D}=\half$;
it is found in the helicity-preserving polarization channel. 
The smallest average visibility is also found in this channel, its value is
$\mathcal{V}=\frac{2}{5}$. 
Even if a direct quantitative comparison with the real CBS
situation cannot be made at this stage, this result is nevertheless consistent
with the experimental observation that the lowest CBS peaks are
actually found in this detection channel \cite{Labeyrie99}.

In summary, we have demonstrated that the concept of wave-particle
duality proves relevant and useful for our understanding of some aspects 
of the interference effects in multiple scattering. 
To make solid quantitative contact with actual CBS experiments, 
the analysis must be extended to account for scattering by three and 
more atoms. 
The stage for this future research is now set.

Ch.~M. and C.~M. wish to thank the Science Faculty 
and the Physics Department of NUS for their kind hospitality and
financial support.
This work was supported by by A$^*$Star Grant No.\ 012-104-0040,
and by NUS Grant WBS: R-144-000-179-112.


\begin{thebibliography} {99}
\bibitem{threehalf}
In the case of a $\half\leftrightarrow\frac{3}{2}$ transition,  
$T=(\boldsymbol{\sigma} \,\boldsymbol{\sigma})-3\mathbf{1}$ has an admixture
of the unit dyadic \cite{HYP}.
\end{thebibliography}
\end{document}